\documentstyle[11pt]{article}
\begin{document}
\baselineskip 22pt 

\pagestyle{empty}
\rightline{CU-TP-804} 
\rightline{SNUTP 96-124}
\rightline{hep-th/9612183}
\rightline{\today} \vskip1cm
\centerline{\Large\bf The BPS Domain Wall Solutions}
\centerline{\Large\bf in Self-Dual Chern-Simons-Higgs Systems}
\vskip 1.5cm
\centerline{\large\it Hsien-chung Kao\footnote{electronic mail:
hckao@mail.tku.edu.tw}}
\vskip2mm
\centerline{Department of Physics, Tamkang University, Tamsui, Taiwan 25137, 
R.O.C.}
\vskip3mm

\centerline{\large\it Kimyeong Lee\footnote{electronic mail:
klee@phys.columbia.edu}}

\vskip2mm
\centerline{Department of Physics, Columbia University, New York, NY
10027, USA}
\vskip3mm

\centerline{\large\it Taejin Lee\footnote{electronic mail:
taejin@cc.akngwon.ac.kr}} 
\vskip2mm 
\centerline{Department of Physics,
Kangwon National University, Chuncheon 200-701, Korea}
\vskip5mm
\vskip10mm
\centerline{\bf ABSTRACT}
\vskip 5mm
\begin{quote}
We study domain wall solitons in the relativistic self-dual
Chern-Simons Higgs systems by the dimensional reduction method to two
dimensional spacetime. The Bogomolny bound on the energy is given by
two conserved quantities in a similar way that the energy bound for
BPS dyons is set in some Yang-Mills-Higgs systems in four dimensions.
We find the explicit soliton configurations which saturate the energy
bound and their nonrelativistic counter parts.  We also discuss the
underlying $N=2$ supersymmetry.
\end{quote}

\noindent{PACS number: 11.27.+d, 11.10.Kk, 11.30.Pb}
\newpage
\pagestyle{plain}
\setcounter{page}{1}
\setcounter{footnote}{0}

\vskip 5mm

\section{Introduction}

In past years the Abelian self-dual Chern-Simons Higgs system with an
appropriate potential has been studied
extensively~\cite{hong,jack1,kim1}. One of its interesting properties
is that there is a Bogomolny-type bound on the energy~\cite{bogo},
which is saturated by the self-dual solitons which carry the
fractional spin and satisfy the fractional statistics. The potential
of this model has two degenerate vacua, implying the existence of
topological domain walls interpolating between them.  It turns out
that the topological domain walls satisfy another Bogomolny bound as a
two dimensional model, which is different from such a bound on three
dimensional dimensional self-dual solitons~\cite{jack1}.

These domain walls which interpolate between the symmetric and
asymmetric phases have been studied to understand the behavior of the
rotationally symmetric solitons in large charge limit. In this case
the energy density is concentrated on a circular ring of large radius,
and its radial cross section resembles the domain
wall~\cite{jack1}. However in the symmetric phase there exist also
rotationally symmetric nontopological solitons with vortices in the
center, whose large charge and vorticity limit cannot be described
exactly by the topological domain walls. This suggests us a richer
structure of domain walls.

On the other hand, recently some two-dimensional non-linear sigma models
with appropriate potentials~\cite{townsend} have been shown to have
many properties similar to the 4-dimensional Yang-Mills Higgs systems
which admit the BPS dyons solutions. In particular, a Noether charge
and a topological charge are found to set the BPS-type bound on the
energy~\cite{bogo,prasad}.  Since these two-dimensional self-dual
models have the underlying $N=2$ supersymmetry and so is our
Chern-Simons-Higgs systems, our domain walls are also expected to have
the BPS bound, which is more general than the Bogomolny bound.

The goal of this paper is to explore the domain wall structure of the
self-dual Chern-Simons-Higgs system, by taking the dimensional
reduction of the model to two dimensions.  We show that the resultant
two dimensional model has a BPS-like energy bound. Our model hence
forms a new class of two dimensional models with the BPS-like energy
bound, similar to the four dimensional Yang-Mills Higgs systems. We
show that the domain wall configurations saturating the energy bound
consist of topological and nontopological domain walls: topological
domain walls interpolating the symmetric and asymmetric phases, and
nontopological domain walls residing in the symmetric phase.

The plan of this paper is as follows: In Sec.~2,~we introduce our
model and find the BPS-type energy bound. We reduce the self-dual
equations to a single ordinary nonlinear differential equation. In
Sec.~3, we solve the differential equation to find and investigate all
possible domain wall solutions.  In Sec.~4, we study the underlying
$N=2$ supersymmetry. In Sec.~5, we study the nonrelativistic limit and
its solitons in the symmetric phase. In Sec.~6, we conclude with some
remarks.  In the appendix, we study another bound on the energy
functional, which works only for the topological domain walls, and
discuss its relation to the BPS bound.

\section{Model}

We start with the self-dual Abelian Chern-Simons-Higgs system whose 
Lagrangian~\cite{hong} is
\begin{equation}
 {\cal L}_{CSH} = {\kappa \over 2}\epsilon^{\mu\nu\rho}A_\mu\partial_\nu
A_\rho + |D_\mu\phi|^2 - {1\over \kappa^2}|\phi|^2(|\phi|^2-v^2)^2 ,
\end{equation}
where $D_\mu\phi =(\partial_\mu -iA_\mu)\phi$. There are two phases in
the theory: the symmetric phase where $\phi = 0$ and the antisymmetric
phase where $\phi = v$. Elementary charged excitations in the
symmetric phase has mass $\mu = v^2/\kappa$ and spin $s= 4\pi/\kappa$,
and in the broken phase there are charge neutral particles of mass
$2\mu$.  There are also self-dual anyonic solitons in the symmetric
and asymmetric phases.

We first compactify the $y$ direction on a circle of radius $r$ and
require that all the fields to be periodic under $y\rightarrow y+2\pi
r$. There are topologically distinct large gauge transformations $e^{in
y/r}$ with any integer $n$. When we expand  $A_y(t,x,y)$ in Fourier
series of $y$, the $y$-independent component $N(t,x)$ becomes periodic
due to the large gauge transformations and lies in the interval
$[-1/2r,1/2r]$. In the limit $r=0$, the low energy physics will be
dominated by the modes which are independent of the $y$ coordinate.

After we rename $\sqrt{2\pi r}\phi$ by $\phi$, $\sqrt{2\pi r}v$ by
$v$, and $2\pi r \kappa$ by $\kappa$ and keep only the $y$-independent
modes, we get the dimensionally reduced Lagrangian in 2 dimensional
spacetime,
\begin{equation}
 {\cal L} = \kappa N F_{01} + |D_\mu \phi|^2 -U(N,\phi),
\end{equation}
where the potential $U(N,\phi)$ is 
\begin{equation}
 U(\phi, N) = N^2|\phi|^2 + \frac{1}{\kappa^2}
|\phi|^2(|\phi|^2-v^2)^2.
\end{equation}
The mass dimension of the fields and parameters become as follows:
$[\phi]=[v]=M^0$, $[A_\mu]=[N]=M$ and $[\kappa]=M^{-1}$. The kinetic
part of $N, A_\mu$ is that of the so-called $BF$ theory. To the
original Lagrangian, one can add an $\theta$ term,
\begin{equation}
{\cal L}_\theta  =  \theta F_{01}, 
\end{equation}
which is possible only in two dimensions. It does not play any role
classically and will be neglected in this paper.

The classical vacuum structure of this theory is determined by the
zeroes of the potential $U(\phi,N)$. The symmetric phase $<\phi>=0$ is
infinitely degenerate as $<N>$ can take any value $N_0$.  In the
symmetric phase there are charged scalar bosons of mass
\begin{equation}
m=\sqrt{N_0^2+\mu^2}
\end{equation}
with $\mu=v^2/\kappa$ being the three dimensional mass.  The
asymmetric phase is uniquely given as $<N>=0,<\phi>=v$ up to the gauge
transformation.  In the asymmetric phase, there are two kinds of
neutral bosons of mass $2\mu$. Since all the vacua are degenerate, we
expect there exist topological domain walls. (We call the domain walls
residing in the symmetric phase to be nontopological, even though they
are topological in a way as they interpolate between different $N$
vacua.)

The Gauss law constraint  from the variation of $A_0$ is 
\begin{equation}
\kappa N' -i(D_0\phi^*\phi-\phi^*D_0\phi) = 0.
\label{gauss}
\end{equation}
where the prime is the derivative $d/dx$.
The variation of  $ N$ yields another constraint,
\begin{equation}
\kappa F_{01} -2N|\phi|^2 = 0,
\label{constr}
\end{equation}
which does not play any important role here.

The theory is invariant under the local gauge transformations
\begin{equation}
\delta\phi = i\Lambda(x)\phi,\,\,\,\, \delta A_\mu = \partial_\mu
\Lambda. 
\end{equation}
Its global part is the $U(1)$ symmetry, whose conserved current is 
$j_\mu = i(D_\mu\phi^*\phi - \phi^*D_\mu\phi)$. Making use of the the
Gauss law, we can write the corresponding charge as
\begin{equation}
Q = \int dx \,\, j_0 = \kappa \Delta N,
\end{equation}
where we introduce  $N_\pm =N(t,\pm \infty)$, the constant asymptotic
values, and define  
their difference as $\Delta N = (N_+ - N_-)$. $Q/\kappa$ can be
identified by the magnetic flux per unit length in three dimensions.
There is also a conserved charge, corresponding to the translation
invariance along the $y$ axis before the dimensional reduction,
\begin{eqnarray} 
 P_y  & = &  \int dx  \,\, i(D_0\phi^* \phi -  \phi^*D_0\phi) N
\nonumber\\
& = &  \frac{\kappa}{2} (N_+^2- N_-^2) =Q \bar{N},
\end{eqnarray}
where we used the Gauss law and  the average of $N$ is defined  as
\begin{equation}
\bar{N}=(N_+ + N_-)/2.
\end{equation}
Thus the $\bar{N}$ can be regarded as the momentum carried by the
unit charge along the $y$ direction. If we identify $\bar{N}=N_0$, the
mass $m=\sqrt{N_0^2+\mu^2} $ of elementary charged particles in the
symmetric phase can be thought to be a Lorentz-boosted  one from three
dimensional mass $\mu$ by the $y$ momentum $N_0$.

For the conserved $P_y$, there exits the corresponding transformation
of the fields,
\begin{eqnarray}
\delta N=0, \,\, \delta \phi = iN\phi, \,\, \delta A_1 = j_0 /\kappa, 
\,\, \delta A_0 = j_1/\kappa.
\end{eqnarray}
This transformation is somewhat puzzling as it does not leave the
Lagrangian invariant without the gauge field equation, although the
Hamiltonian is invariant modulo the Gauss law.  Even though the $y$
space has disappeared, there remains this $y$ directional translation.

Finding the BPS energy bound begins with the explicit expression of
the energy density, 
\begin{equation}
{\cal E} = |D_0\phi|^2 +|D_1\phi|^2+N^2|\phi|^2 +
\frac{1}{\kappa^2}|\phi|^2(|\phi|^2-v^2)^2. \label{energy}
\end{equation}
Together with the Gauss law, one can put the energy density as
\begin{eqnarray}
{\cal E} = & & |D_0 \phi +  \frac{i}{\kappa}\phi (|\phi|^2-v^2) \sin
\alpha -   iN \phi \cos\alpha |^2 \nonumber \\
& & + |D_1\phi + \frac{1}{\kappa} \phi(|\phi|^2-v^2) \cos\alpha
 +  N \phi \sin\alpha |^2  \nonumber \\
& &   +  \frac{1}{2\kappa} \biggl[ \kappa^2 N^2 - (|\phi|^2-v^2)^2
\biggr]' \cos\alpha  + [N(v^2-|\phi|^2)]' \sin\alpha. \label{energyb}
\end{eqnarray}
where $\alpha$ is an arbitrary angle variable. We introduce two
charges, 
\begin{eqnarray}
& &  Y = \int dx   \frac{1}{2\kappa} \biggl[
\kappa^2 N^2 - (|\phi|^2-v^2)^2 \biggr]', \label{yeq}\\
& &  Z = \int dx \biggl[N(v^2  -|\phi|^2 )\biggr]'. \label{zeq}
\end{eqnarray}
The boundary conditions at spatial infinities for any finite
configuration should approach the ground configuration of the
potential, $(F_\pm=1,N_\pm=0)$ or $(F_\pm=0,N_\pm={\rm any}\,\,{\rm
value})$, where $|\phi|^2=v^2 F$. We can rewrite the above two charges
as
\begin{eqnarray}
& &  Y = P_y -\frac{1}{2} \kappa \mu^2 (F_+-F_-)(F_+ + F_- -2),  \\
& & Z =  \mu Q. 
\end{eqnarray}
Thus we can identify $Y$ as the topological charge and $Z$ as the
Noether charge in a broad sense.

{}From Eq.~(\ref{energyb}), we can now put a bound on the energy for
configurations of a given $Y$  
and $Z$ as
\begin{equation}
 E \ge   Y \cos\alpha + Z \sin{\alpha} \label{eb1}
\end{equation}
for any  $\alpha$. It follows that
\begin{equation}
E \ge \sqrt{Y^2 + Z^2}. 
\end{equation}
This is the BPS-like bound. The bound is saturated by the field
configurations satisfying the constraint
equations  (\ref{gauss}) and (\ref{constr}), and the following
self-dual equations: 
\begin{eqnarray}
& & D_0 \phi +  \frac{i}{\kappa}\phi (|\phi|^2-v^2) \sin
\alpha  -   iN \phi \cos\alpha =0, \label{self1}\\
& & D_1\phi + \frac{1}{\kappa}  \phi(|\phi|^2-v^2) \cos\alpha
 +  N \phi \sin\alpha =0. \label{self2}
\end{eqnarray}
Here $\cos\alpha=Y/\sqrt{Y^2+Z^2}$ and $\sin\alpha =
Z/\sqrt{Y^2+Z^2}$.  The self-dual configurations are static in time as
one can show easily that $\partial_0 |\phi|^2 = \partial_0 N =
\partial_x \,{\rm Arg} \phi - A_x = 0 $. Furthermore, the self-dual
configurations also satisfy the usual second-order field equations.
The additional constraint (\ref{constr}) is automatically satisfied by
the field configurations satisfying the self-dual equations and the
Gauss law.

The dimensional reduction of the Bogomolny-type energy bound in three
dimensions is identical to the above case with $\sin\alpha = \pm 1$
and so is not a new bound.  However in the appendix we will
describe another Bogomolny-type bound which works only for the
topological domain walls.

Again with $F=|\phi|^2/v^2$, we can combine the Gauss law
(\ref{gauss}), and Eqs.~(\ref{self1}) and (\ref{self2}) as 
\begin{eqnarray}
& &  N'= -2\mu F \bigl\{  \mu(F-1)\sin \alpha -N  \cos\alpha
\bigr\}, \label{sd1} \\ 
& &  F'= - 2F\bigl\{ \mu(F-1)\cos\alpha + N\sin\alpha \bigr\}, 
\label{sd2}
\end{eqnarray}
where the prime is $d/dx$.  The above equations can be put together as a
single ordinary differential equation, 
\begin{eqnarray}
(\ln F)'' & = & -2(\mu F'\cos\alpha + N'\sin\alpha) \nonumber\\
& = & - 4\mu^2 F(1-F).
\end{eqnarray}
After integrating this equation once, we obtain 
\begin{equation} 
F'^2-4\mu^2F^2(F^2-2F+a^2) = 0,     \label{feq}
\end{equation}
where $a$ is an integration constant that lies in $[0, 1]$ for any
finite energy density solution. For a given $F$ configuration
Eq.~(\ref{sd2}) determines $N$.

There is one different feature in our model compared with the
previously studied BPS-type solitons: The length scale of the above
equation (\ref{feq}) is independent of the parameter $\alpha$,  and so
is that of the self-dual domain walls.

\section{Self-dual Solitons}

\subsection{between the symmetric and asymmetric phases}

For  $a=1$, the solutions of Eq.(\ref{feq}) are 
\begin{equation}
 F= {1\over 1+ e^{\mp 2\mu x }} \label{topd1} ,
\end{equation}
satisfying $F' = \pm 2 \mu F(1-F)$. These solutions describe
topological domain walls interpolating between the symmetric and
asymmetric phases. The transition region from the symmetric phase to
the asymmetric one has the size of order $1/\mu$.  The scalar field $N$
from Eq. (\ref{sd2}) is
\begin{equation}
N = \frac{\mu(\mp 1 + \cos \alpha)}{\sin\alpha \,\,(1 +e^{\pm 2\mu x}) 
}. \label{topd2}
\end{equation}
Of course the position of the solitons can be translation  in space by
replacing $x$ by $x-c$. This is the only zero mode of the solution.

The solutions can be classified into two classes depending whether
$F_+$ is one or zero.  The first case with $(F_+,F_-) = (1,0)$,  we
get $N_+=0$ and $N_-=\mu(-1+\cos\alpha)/\sin\alpha$, which fixes
$\alpha$ in terms of $N_-$. The total energy is
\begin{equation}
 E = \frac{\kappa \mu^2 (1-\cos\alpha)}{\sin^2\alpha}. \label{dome}
\end{equation}
The topological charges for this soliton are $Y=E\cos\alpha$ and
 $Z=E\sin\alpha$. In addition, $Q=-\kappa N_-$ and $P_y=-\kappa
 N_-^2/2$. Near $\alpha\approx 0$, the wall does not carry any charge
 and the energy takes $E=\kappa \mu^2/2$, the minimum value of
 Eq.~(\ref{dome}).

The second case is with $(F_+,F_-) = (0,1)$, which is the spatial
reflection of the first solution. In this case
$N_+=\mu(1+\cos\alpha)/\sin\alpha$, which fixes the $\alpha$ in terms
of $N_+$, and $ N_-=0$.  The energy of the soliton is
\begin{equation}
E = \frac{\kappa \mu^2 (1+\cos\alpha)}{\sin^2\alpha}. \\
\end{equation}
Again the charges are uniquely determined by $N_+$ or the angle $\alpha$.

\subsection{In the symmetric phase}

For $a< 1$, the general solution of Eq.(\ref{feq}) is given by
\begin{equation}
 F = \frac{a^2}{ 1+ \sqrt{1-a^2} \cosh(2a\mu x) }.
\label{relsymm}
\end{equation}
Since $F_\pm=0$, this domain wall lives in the symmetric phase.
{}From Eq.(\ref{sd2}), we get
\begin{equation}
N = \biggl( \frac{\mu\sqrt{1-a^2}}{\sin\alpha} \biggr)
\biggl( \frac{ a \sinh(2a\mu x)+ [\sqrt{1-a^2}+ \cosh(2a\mu x)]
\cos\alpha }{1+ \sqrt{1-a^2} \cosh (2a\mu x)} \biggr),
\end{equation}
where $N_+ = \mu(a+\cos\alpha)/ \sin \alpha$ and
$N_-=-\mu(a-\cos\alpha)/\sin\alpha$.
The energy is given by
\begin{equation}
E = \frac{2 a\kappa \mu^2}{ \sin^2 \alpha}.
\label{en1}
\end{equation}
Thus, for  given $\alpha$ and $a$, there is a unique solution up to the
translation. Again, $Y$ and $Z$ are $E\cos\alpha$ and $E\sin\alpha$.

Let us reconsider the above configuration somewhat differently.  The
average of $N$ is $\bar{N}=\mu\cot\alpha$. This fixes the angle
$\alpha$. The difference of $N$ is given by $\Delta N =2\mu
a/\sin\alpha$, which fixes $a$ in terms of $\bar{N}$ and $\Delta N$.
Since the electric charge is $Q=\kappa \Delta N$, the  energy
(\ref{en1}) can be expressed as
\begin{equation}
E = Q \sqrt{\bar{N}^2 + \mu^2}.
\end{equation}
Since the mass of elementary charged particles in the symmetric phase
is $m=\sqrt{N_0^2+\mu^2}$, this soliton  can be regarded as a
Q-ball lump of  elementary particles if we identify the vacuum
value $N_0$ by the average $\bar{N}$.

In the limit $a\rightarrow 1$, the size of this soliton gets very
large. The spatial dependence is characterized by the parameter
$\mu$. As $x$ increases, the value of $F$ jumps from zero to a value
approximately one in a wall of size $1/\mu$, then stays there for the
interval of approximate size $(-\ln\sqrt{1-a^2})/2\mu$, then falls
down to zero in the wall of size $1/\mu$. {}From the spatial
dependence of $F$ and $N$, one can see that in this limit the soliton
looks more and more like a combination of two topological solitons
considered in the previous subsection.

Especially when we choose $\bar{N}=0$ or $\cos \alpha=0$, the limit
$Q= 2\kappa\mu a  \rightarrow 2v^2$ as $a\rightarrow 1$. If
the semiclassical picture should be correct, the charge of this
soliton should be much larger than that of elementary particles, or
$2v^2 \gg 1$.

\section{Supersymmetry}

In Ref.~\cite{choonkyu} the underlying $N=2$ supersymmetry theory for
the self-dual Chern-Simons Higgs systems has been found. After the
dimensional reduction of this model, we get again an $N=2$
supersymmetric model.  Here we use the convention that $\eta_{\mu\nu}
= {\rm diag}(1, -1, -1),$ and $\gamma^0 = \sigma_2, \gamma^1 =
i\sigma_1, \gamma^2 = -i\sigma_3.$ After the dimensional reduction, we
introduce $\gamma^5=\gamma_5 \equiv \gamma^0 \gamma^1=\sigma_3$.  The
supersymmetric Lagrangian in two dimensions becomes
\begin{eqnarray}
{\cal L}_{susy} & = & \kappa N F_{01} + |D_\mu \phi|^2 - N^2 |\phi|^2
- {1\over \kappa^2} |\phi|^2(|\phi|^2 -v^2)^2 \nonumber \\
&\,& + i\bar{\psi}\gamma^\mu D_\mu \psi - iN \bar{\psi}\gamma^5 \psi 
+ {1\over \kappa} (3|\phi|^2 -v^2) \bar{\psi}\psi. 
\end{eqnarray}
After the dimensional reduction, the supersymmetric transformation becomes
\begin{eqnarray}
& & \delta  A_\mu  =  {1\over \kappa}(\bar{\zeta}\gamma_\mu \psi \phi^*
+ \bar{\psi}\gamma_\mu \zeta \phi), \nonumber\\
& & \delta N  =  {i\over \kappa}(\bar{\zeta}\gamma^5 \psi \phi^*
+ \bar{\psi}\gamma^5 \zeta \phi), \nonumber\\
& & \delta \phi  =  \bar{\zeta} \psi, \\
& & \delta \psi  = - i\gamma^\mu\zeta D_\mu \phi  +i\gamma^5\zeta N
\phi +{1\over \kappa}\phi( |\phi|^2 -v^2)\zeta. \nonumber
\end{eqnarray}
Since the parameter $\zeta$ is a complex Dirac spinor, it generates
the $N=2$ supersymmetry transformation. (We can choose the chiral
spinors $(1\pm \gamma^5)\zeta/2$ as independent parameters  and so the
supersymmetry is really $N=(2,2)$.)

For a given superfield $\Phi$ the supersymmetric transformation 
\begin{equation}
\delta \Phi = [i(\bar{\zeta} {\cal R} + \bar{\cal R} \zeta), \Phi],
\end{equation}
is generated by the supercharge
\begin{equation}
{\cal R} = \int dx \biggl\{ \bigl[ (D_\mu \phi)^* \gamma^\mu
+ N\phi^* \gamma^5- {i\over \kappa} \phi^*(|\phi|^2 -v^2)\bigr]
\gamma^0 \psi \biggr\}.
\end{equation}

After quantization, we get the canonical commutation relations between
the fields. These lead to the  relevant $N=2$  superalgebra,
\begin{equation}
\left[\bar{\zeta}{\cal R}, \bar{\cal R} \eta \right]
= \bar{\zeta} \gamma^\mu \eta P_\mu - \bar{\zeta} \eta Z +
i\bar{\zeta} \gamma^5 \eta Y, \label{super} 
\end{equation}
where two central charges are given as 
\begin{eqnarray}
& & Y = \int dx \left\{ N \bigg[ i (D_0\phi^*\phi - \phi^*D_0\phi ) -
\psi^\dagger\psi\bigg] - \frac{1}{2\kappa} \partial_x (|\phi|^2-v^2)^2
\right\},\\
& & Z = -\int dx \left\{ N \partial_x |\phi|^2 +
\frac{1}{\kappa}(|\phi|^2-v^2) \biggl[ i (D_0\phi^*\phi-\phi^*D_0\phi)
-\psi^\dagger\psi )\biggr] \right\}.
\end{eqnarray} 
These central charges become those in Eqs.~(\ref{yeq}) and (\ref{zeq})
once the Gauss law (\ref{gauss}) is used.

After introducing a new spinor  operator,
\begin{eqnarray}
{\cal R}' & = & \frac{1}{2} \left( 1 + \gamma_0 e^{i(\pi/2-
\alpha)\gamma_5}\right) {\cal R} \nonumber \\ 
& = & \frac{1}{2}\left( 1 + \gamma_0 e^{i(\pi/2-\alpha)\gamma_5}
\right) \int dx  \biggl\{ \psi
\left[ (D_0 \phi)^* - \frac{i}{\kappa} \phi^*(|\phi|^2 - v^2)  \sin\alpha
+  iN\phi^* \cos\alpha \right] \biggr. \nonumber\\
& \ &\qquad\qquad\qquad\qquad\qquad\;\,  \biggl. - \gamma_5 \psi
\left[ (D_1 \phi)^*  + \frac{1}{\kappa} \phi^*(|\phi|^2 - v^2)  \cos\alpha
+N\phi^*  \sin\alpha  \right]\biggr\}, 
\end{eqnarray}
the superalgebra (\ref{super}) becomes 
\begin{equation}
\sum_\beta \left\{ {\cal R}'_\beta, {\cal R}_\beta'^\dagger \right\}
= E - \left(  Y \cos\alpha  + Z\sin\alpha  \right).
\end{equation}
As the left-handed side of this equation is positive definite, this
equation leads to the energy bound (\ref{eb1}).  The energy bound is
saturated when $<{\cal R}'> = 0$, which in turn implies the self-dual
equations.

\section{The Nonrelativistic Limit}

We expect the nonrelativistic limit of the theory can be taken in the
symmetric phase as the field $\phi$ describes charged particles.  We
expect the nonrelativistic limit is reasonable if the kinetic energy
is small. For this limit to make a sense, it turns out that the $y$
momentum per unit charge should be also small compare with the
three-dimensional mass, or $|N_0|\ll \mu$.

In this case we take the non-relativistic limit of the system in the
three dimensional model by letting $\phi = (1/\sqrt{2\mu})e^{-i\mu
t}\psi,$ where $\mu =v^2/\kappa$. The Lagrangian (1) reduces to the
well known Lagrangian~\cite{pi},
\begin{equation}
{\cal L}_{nonrel} = {\kappa \over
2}\epsilon^{\mu\nu\rho}A_\mu\partial_\nu A_\rho + i\psi^* D_t \psi -
{1\over 2\mu}|D_\mu\psi|^2 + {1\over 2\mu\kappa}(|\psi|^2)^2 .
\end{equation}
After dimensional reduction, it becomes
\begin{equation}
{\cal L}_{nonrel} =\kappa N F_{01}+ i\psi^* D_t \psi - {1\over
2\mu}|D_x\psi|^2 -{1\over 2\mu}N^2|\psi|^2 + {1\over
2\mu \kappa}(|\psi|^2)^2.  
\end{equation}
This is equivalent to the nonrelativistic limit of the dimensionally
reduced Lagrangian if $N_0 \ll\mu$.
The energy functional, after dropping the boundary term which depends
only on the conserved charge, is given by
\begin{equation}
E_{nonrel} =\int dx \left\{{1\over 2\mu}|D_x\psi|^2 +{1\over
2\mu}N^2|\psi|^2 - {1\over 2\mu \kappa}(|\psi|^2)^2 \right\}. 
\end{equation}
This energy is reasonable as its density is gauge invariant. We split
the $N$ field as a sum of the average $\bar{N}$ and the fluctuation
$\delta N$ such that $\delta N_+ + \delta N_-=0$.

By making use of the Gauss law $\kappa N' + |\psi|^2 = 0$, the energy
functional can be rewritten as 
\begin{equation}
E_{nonrel} = \frac{1}{2\mu}\bar{N}^2 |Q| +  \int dx \left\{{1\over
2\mu}|D_x\psi - \delta N\psi|^2  \right\},  
\end{equation}
plus the vanishing boundary terms, where $Q=-\int dx |\psi|^2<0$.
Since the integrand is nonnegative, the nonrelativistic energy is
bounded by the $y$ directional kinetic energy, $\bar{N}^2|Q|/(2\mu)$,
which can be seen as the mass correction of the $Q$ particles by
$P_y=\bar{N}$.

Introducing $|\psi|^2 = 2\mu v^2 F,$ we can simplify the Gauss law
and the nonrelativistic  self-dual equations to
\begin{eqnarray}
& & \delta N' +2\mu^2 F = 0, \\
& & F' - 2 \delta NF = 0.
\end{eqnarray}
They in turn lead to the following equations
\begin{eqnarray}
(\ln F) '' + 4\mu^2 F = 0, \label{liou} \\
\delta N= (\ln F)'/2.
\end{eqnarray}
The equation (\ref{liou}) is the Liouville equation in one dimension,
which can be integrated to 
\begin{equation}
F'^2  +4\mu^2 F^2(2F - a^2) = 0,
\end{equation}
where $0<a<1$. This is again the  nonrelativistic limit of
Eq.~(\ref{feq}) with the same parameter $a$.
The solutions are given by
\begin{eqnarray}
& & F = {a^2 \over 2 \cosh^2(a\mu x)}, \\
& & N = N_0-\frac{a\mu}{2} \tanh(a\mu x). 
\end{eqnarray}
In the large $x$ region the above nonrelativistic solutions match with
the relativistic solution (\ref{relsymm}) if $a \ll 1, \sin \alpha = -1$. 
The charge of this configuration is again  $Q=2\kappa\mu a$. 

\section{Concluding Remarks}

Here we have explored the structure of the domain walls in the
self-dual Chern-Simons-Higgs systems by the dimensional reduction
method. Their energy bound is similar to the BPS bound on dyons in
some Yang-Mills Higgs systems in four dimensions. We found all
possible domain solutions, which are made of topological and
nontopological domain walls. We studied the $N=2$ supersymmetry behind
the BPS-like energy bound. We also have studied the nonrelativistic
limit.

Our two dimensional model is different from the recent attempts to
describe anyons in one dimensional space~\cite{rabello}. Neither our
model seems to be directly related to the Calogero-Suthutherland
model, contrast to the recent work~\cite{andric}. Our nonrelativistic
limit is simpler than the recently discovered chiral
solitons~\cite{agli}.  However, it turns out that our
nonrelativistic model has also intriguing properties as shown in a
recent work~\cite{poh}.

There are several directions to explore from this point.  There are
self-dual Chern-Simons-Higgs models for any gauge group and any matter
representation~\cite{klee1,dunne1}. Especially with the matter in the
adjoint representation, the vacuum structure is quite
rich~\cite{kao1,dunne2}, implying the intricate domain wall
structure. As in the abelian case, domain walls in two dimensions are
generally easier to understand than solitons in three dimensions, and
such understanding will eventually lead to a better understanding of
solitons in three dimensions.

We studied the $N=2$ supersymmetry underlying the self-duality.
However, in three dimensions with the Chern-Simons term the maximal
supersymmetry is $N=3$~\cite{kao3}, which would be translated to the
$N=3$ supersymmetric theory in two dimensions. As a pure two
dimensional system, it is interesting to find out how the $N=3$
supersymmetry can be maximal. One may wonder whether there exist a
more general set of the self-dual systems with the $BF$ kinetic termin
two dimensions with larger supersymmetry than the $N=3$ supersymmetry.

Along the similar line of thought, there may be a gauged version of
the nonlinear sigma models considered in Ref.~\cite{townsend}.  They
would lead to a richer variety of two dimensional models with the
BPS-type energy bound, which may be similar to the dimensional
reduction of the self-dual $CP(N)$ models considered in three
dimensions~\cite{kimm}.

The quantum mechanical aspect of the theory should be also
interesting. In the massless limit $v=0$, there may be a quantum
conformal symmetry.  Also the three dimensional solitons may appear as
instantons in two dimensions, whose effect is unclear at this
moment. One of the noticeable feature here is that the Euclidean
action is complex and so the instanton solutions should be treated
carefully as in the case of monopole instantons in three dimensional
Chern-Simons-Higgs systems~\cite{klee4}.

\vskip1cm

\noindent{\bf Acknowledgements}

The work of H.-C. K. was supported by the National Science Council of
Taiwan-Republic of China under the contract number No. NSC
86-2112-M-032-011-T. The work of K.L. was supported by the US
Department of Energy and the NSF Presidential Young Investigator
Fellowship. The work of T.L. was supported by the Basic Research
Institute Program, Ministry of Education, Project BSRI-96-2401 and in
part by KOSEP through the Center for Theoretical Physics of Seoul
National University.

\newpage
\appendix
\noindent{\bf Appendix: Another bound}

In this appendix, we briefly review another Bogomolny-type bound which
works only for the topological domain walls~\cite{jack1}.
We can rewrite the energy (\ref{energy}) as follows:
\begin{eqnarray}
{\cal E} = & & |D_0 \phi \pm iN \phi |^2 + |D_1\phi \pm
\frac{1}{\kappa} \phi(|\phi|^2-v^2) |^2 \nonumber 
\\ 
& & \mp \frac{1}{2\kappa} \biggl[ \kappa^2 N^2 + (|\phi|^2-v^2)^2 \biggr]'.
\label{energyc}
\end{eqnarray}
This is not identical to the equation (\ref{energyb}) with
$\cos\alpha=\pm1 $ because of the sign difference in the boundary
term.  The energy bound is then
\begin{equation}
E \ge \biggl| \int dx \frac{1}{2\kappa} \bigl[ \kappa^2 N^2 +
(|\phi|^2-v^2)^2 \bigr]' \biggr| \label{append}
\end{equation}

Let us concentrate on the upper sign.
The self-dual equations in this case become
\begin{eqnarray}
& & \kappa N'= -2|\phi|^2 \label{app1} \\
& & D_1\phi + \frac{1}{\kappa}\phi(|\phi|^2-v^2) = 0 \label{app2}
\end{eqnarray}
where we have used the Gauss law (\ref{gauss}).  The $\phi$ equation
leads to the topological domain wall solution (\ref{topd1}) with
$(F_+,F_-)= (1,0)$.  The $N$ equation has a nontrivial solution
(\ref{topd2}) with $(N_+=0,N_->0)$.

Seemingly different two sets of self-dual equations are satisfied by
the fields for the same topological domain wall. The new energy bound
(\ref{append}) is identical to the BPS energy (\ref{dome}) for these
configurations. We believe that the presence of this additional energy
bound is due to the interdependence of two charges $Y$ and $Z$ for the
topological domain walls. While the above bound does not lead to
anything new in the abelian self-dual case, its analogue in the
nonabelian case may be more useful.


\newpage

\begin{thebibliography}{}

\bibitem{hong}
{J. Hong, Y. Kim and P-Y. Pac,   Phys. Rev.
Lett.   64, 2330 (1990); R. Jackiw and E. Weinberg, Phys.
Rev. Lett.  64, 2334 (1990).}

\bibitem{jack1}{R. Jackiw, K. Lee and E.J. Weinberg,  Phys. Rev.  D42,
 3488 (1990). }

\bibitem{kim1}{Y. Kim and K. Lee,  Phys. Rev.  D49, 2041  (1994).}

\bibitem{bogo}
{E. Bogomolny,  Sov. J. Nucl. Phys.  24, 449 (1976).}


\bibitem{townsend}{E.R.C. Abrahm and P.K. Townsend, Nucl. Phys.  B 351, 
313 (1991); Phys. Lett. B291, 85 (1992);
Phys. Lett.  B295, 225 (1992); A. Opfermann and G. Papadopoulos, 
{\it Solitons in (1,1)-supersymmetric massive sigma models},
hep-th/9610099.}


\bibitem{prasad}{M.K. Prasad and C.M. Sommerfield,
Phys. Rev. Lett.  35, 760 (1975); S. Coleman, S. Parke, A. Neveu
and C.M. Sommerfield, Phys. Rev. D15, 544 (1977).}

\bibitem{choonkyu}{C. Lee, K. Lee and E. Weinberg, Phys. Lett. B 243, 105
(1990).} 
\bibitem{pi}{R. Jackiw and S.-P. Pi, Phys. Rev. Lett. 64, 2969 (1990);
Phys. Rev. D 42, 3500 (1990).}

\bibitem{rabello}{S.J. Rabello, Phys. Lett. B363, 180 (1995)}


\bibitem{andric}{I. Andri\'c, V. Bardek and L. Jonke, {\it
Calogero-Sutherland model from excitations of Chern-Simons vortices},
hep-th/9507110.}

\bibitem{agli}{U. Aglietti, L. Griguolo, R. Jackiw, S.-Y. Pi, and
D. Seminara, Phys. Rev. Lett. 77, 4406 (1996); R. Jackiw, {\it A
Nonrelativistic Chiral Soliton in One Dimension}, hep-th/9611185}.

\bibitem{poh}{P. Oh and C. Rim, {\it Solitons in the Chern-Simons
Inspired 1+1 D Field Theory}, hep-th/9612028.}

\bibitem{klee1}
{K. Lee, Phys. Lett. {\bf B255}, 381 (1991); 
K. Lee,  Phys. Rev. Lett.  {\bf 66}, 553 (1990).}

\bibitem{dunne1}
{G. Dunne,  Phys. Lett. B324, 359 (1994). }

\bibitem{kao1}
{H-C. Kao and K. Lee,  Phys. Rev. D50, 6626 (1994). }

\bibitem{dunne2}
{G. Dunne,  Nucl. Phys.  B433, 333 (1995); G. Dunne,
 Phys. Lett.  B345, 452  (1995).}

\bibitem{kimm}
{K. Kimm, K. Lee and T. Lee, Phys. Rev. D53, 4436 (1995).} 






\bibitem{kao3}
{H.-C. Kao and K. Lee,  Phys. Rev.  D46, 4691  (1992);
H.-C. Kao,  Phys. Rev.  D50,  2881 (1994).} 


\bibitem{klee4}
{K. Lee,  Nucl. Phys.  B373, 735 (1992).}

\end{thebibliography}
\end{document}